\documentclass[10pt,a4paper]{article}
\usepackage{bm}
\usepackage{graphicx}
\usepackage{amsmath}
\usepackage{amssymb}
\usepackage{cite}
\usepackage{subfig}
\usepackage[top=1in,bottom=1in,right=1in,left=1in]{geometry}

\title{Effect of Confinement of Gluons and One Pion Exchange in Nucleon-Nucleon Interaction}
\author{V K Nilakanthan\footnote{Corresponding author, E-mail:nveluthat@gmail.com} $^{1,2}$, V C Shastry$^{1}$, S Raghavendra$^{1,3}$, and K B Vijaya Kumar$^{1}$}
\date{}
\begin{document}
\maketitle
\begin{center}
\vspace*{-0.25in}
$^{1}$Department of Studies in Physics, Mangalore University, Mangalore, India - 574199.\\
$^{2}$Department of Post Graduate Studies and Research in Physics, St Aloysius College (Autonomous), Mangalore, India - 575003.\\
$^{3}$Department of Post Graduate Studies in Physics, S.D.M. College (Autonomous), Ujire, India - 574240.\\
\end{center}
\begin{abstract}
The Nucleon-Nucleon interaction in the singlet ($^{1}S_{0}$) and triplet ($^{3}S_{1}$) channels have been studied in the framework of Relativistic Harmonic Model using the Resonating Group Method in Born-Oppenheimer approximation. The full Hamiltonian consists of the kinetic energy, two body confinement potential, confined one gluon exchange potential and one pion exchange potential. Contribution of Confined One Gluon Exchange Potential and One Pion Exchange Potential to the adiabatic nucleon-nucleon interaction potential is discussed.
\end{abstract}

Keywords: Nucleon-Nucleon Interaction, Relativistic Harmonic Model, Confined One
Gluon Exchange, One Pion Exchange, Resonating Group Method.

PACS: 13.75.Cs, 12.39.Ki, 21.45.Bc

\section{Introduction}
The nucleon-nucleon (NN) interaction has always been an important and never ending challenge. It is known that nucleus is treated as a system of nucleons and that nucleons are made up of quarks, bound by the strong interaction. Quantum Chromodynamics (QCD) is the widely accepted theory for strong interaction. The two key features in the QCD are color confinement and asymptotic freedom. Since the exact form of the confinement is not known from QCD, phenomenological models have to be employed to study NN interaction. These phenomenological models include either relativistic models \cite{1,2,3,4,5,6} where the interaction is treated perturbatively or non relativistic quark models (NRQM) \cite{7,8,9,10,11,12,13,14} where a suitable potential is used. 

In this work we have  investigated the singlet and triplet NN potential in the frame work of relativistic harmonic model (RHM).  In an earlier work, the NN scattering potentials and phase shifts were obtained without taking into account the contribution of one pion exchange potential (OPEP) \cite{15,16}. The aim was to test the effect of confined gluons on NN scattering phase shifts. Since it is well established that the OPEP provides state-independent repulsion and contributes substantially to the NN adiabatic potential \cite{17,18}, we feel any reasonable quark model should include OPEP to  obtain the partially conserved axial current, consistent with the chiral symmetry, which specifies the coupling of pions with quarks. Hence, a picture of the nucleon of core radii of about 0.6 fm and a pion of small size coupling to quarks would help to clarify why the long-range NN potential could be accurately described by the OPEP, which is of vital importance for nuclear physics.

The pion is treated as an elementary field which couples to the quarks in the framework of NRQM. The strength of the coupling is same as that of the experimental $\pi$-N coupling strength at zero momentum transfer \cite{17}. In an alternative approach, a suitable form factor with a cutoff mass $\Lambda$ is introduced to take care of the effect of the inner structure of pions at short range \cite{19}.

In conventional quark models, the Hamiltonian consists of kinetic energy term, one gluon exchange potential (OGEP) term and confinement potential. In all these models, the exchange part of the color magnetic interaction is considered responsible for the short range repulsion \cite{15,16}. The OGEP is obtained from the QCD Lagrangian in the non relativistic limit by retaining terms to the order of $\frac{1}{c^{2}}$. The gluon propagators used to derive the OGEP are similar to the free photon propagators in QED, which is used to obtain the Fermi-Breit interaction. Since confinement of color implies confinement of quarks and gluons, to determine the nature of NN interaction, a decisive role should be played by the confined dynamics of the gluons. For the confinement of quarks we make use of the relativistic harmonic model (RHM) and for the confinement of gluons we make use of the current confinement model (CCM) \cite{15}. The confined one gluon exchange potential (COGEP) used in the present work is derived in the CCM using confined gluon propagators (CGP) \cite{15,16}. 
 
Recently, nuclear potentials were calculated by lattice QCD by utilizing the long distance behavior of Nambu-Bethe-Salpeter wave functions \cite{20}. Further insights in this work can be seen in References \cite{21,22}. In a work by Vinh Mau et al., it was shown that the behavior of NN interaction at short distances is dependent on additional terms added for the intermediate and long range forces \cite{23}. With the experimental data provided by the J-PARC \cite{24}, PANDA \cite{25}, NICA \cite{26}, and HIAF \cite{27} projects, there is good progress in this regime.

The paper has been divided into 4 sections. The Relativistic Harmonic Model and Resonating group method are discussed in the section 2. The results and discussions are presented in the section 3 and the conclusions in section 4.
    
\section{Model}

\subsection{Relativistic Harmonic Model}

In the RHM \cite{29} the quarks in a nucleon are considered to be confined by a Lorentz scalar plus vector harmonic oscillator potential,
\begin{equation}
\frac{1}{2}(1+\gamma_{0})\alpha^{2}r^{2}+M
\end{equation}
where $\gamma_{0}=\left[
\begin{array}{cc}
1 & 0\\
0 & -1\\
\end{array}
\right]$ is the Dirac's matrix, M is a constant mass and $\alpha^{2}$ is the confinement strength parameter. In RHM the quark wave function $\psi$ is given by, 
\begin{equation}
\psi=N \left[\begin{array}{c}
\phi \\
\frac{{\bm \sigma}.{\bm p}}{E+M}\phi\\
\end{array}
\right]
\end{equation}
where 
\begin{equation}
\nonumber N=\sqrt{\frac{2(E+M)}{3E+M}}
\end{equation}
Here $E$ is an eigen value of the single particle Dirac equation. The lower component of $\psi$ is eliminated to obtain a harmonic oscillator wave equation in $\phi$
\begin{equation}
(\frac{p^{2}}{E+M}+\alpha^{2}r^{2})\phi=(E-M)\phi
\end{equation}

The full Hamiltonian used in this work is
\begin{equation}
H=K+V_{int}+V_{conf}-K_{CM}
\end{equation}
where $K$ is the Kinetic energy, $V_{int}$ is the interaction potential, $V_{conf}$ is the confinement potential and $K_{CM}$ is the kinetic energy of the center of mass. 
\[
K=\sum_{i=1}^{6}\frac{p_{i}^{2}}{(E+M)}
\]
\[
K_{CM}=\frac{P^{2}}{6(E+M)}
\]
where $\frac{(E+M)}{2}$  is the dynamic effective mass of the quarks, $p_{i}$ is momentum of the $i^{th}$ quark and $P$ is the momentum of the center of mass. The interaction potential is given by,
\[
V_{int}=V_{COGEP}+V_{OPEP}
\]
where
\begin{equation}
V_{COGEP}=\frac{\alpha_{s}}{4}N^{4}[D_{0}({\bm r})+\frac{1}{(E+M)^{2}}(4\pi \delta^{3}({\bm r})-c^{2}D_{0}({\bm r}))(1-\frac{2}{3}{\bm \sigma_{i}}.{\bm \sigma_{j}})]{\bm \lambda_{i}}.{\bm \lambda_{j}}
\end{equation}
where 
\[D_{0}(\bm r)=exp(-\frac{r^{2}c_{0}^{2}}{2})(\alpha_{1}\frac{1}{r}+\alpha_{2})
\]
In the above equation, $\bm \lambda_{i}$ and $\bm \lambda_{i}$ are the generators of the color $SU(3)$ group for the $i^{th}$ and the $j^{th}$ quarks, $\bm\sigma_{i}$ and $\bm\sigma_{j}$ are the Pauli spin operators of the $i^{th}$ and the $j^{th}$ quarks and $\alpha_{s}$ is the strong coupling constant.

The OPEP \cite{17} is given by 
\begin{equation}
V_{OPEP}=\frac{f_{q}^{2}}{3}\sum_{i<j}\frac{e^{-m_{\pi}r_{ij}}}{r_{ij}}(\bm\sigma_{i}.\bm\sigma_{j})(\bm\tau_{i}.\bm\tau_{j})
\end{equation}
where $\bm\tau_{i}$ and $\bm\tau_{j}$ are the isospins of the $i^{th}$ and the $j^{th}$ quarks and $f_{q}$ is the OPEP strength parameter.

\subsection{Resonating Group Method}
The NN interaction exists only when there is exchange of quarks between the nucleons. If the quarks are not exchanged between the nucleons, there can be no NN interaction arising from the quark-quark (qq) interaction. This can be seen from the fact that the matrix element of $\bm \lambda_{i}.\bm \lambda_{j}$ vanishes when $i^{th}$ quark is in one nucleon and $j^{th}$ is in the other nucleon in accordance with the Wigner-Eckart theorem. Hence the only way NN interaction can arise from a qq interaction is by constructing a totally antisymmetric wave function for the six quark system in which case the exchange terms arising solely out of anti-symmetrization. The RGM employs a totally antisymmetric wave function and treats the motion of the center of mass correctly. Using the RGM technique we solve the equation \cite{28},
\begin{equation}
\langle \psi | (H-E){\bm A}|\psi\rangle=0
\end{equation}
to get the energy ($E$) of the interacting nucleons. Here, $H$ is the Hamiltonian, $\psi$ is the wave function of the nucleons and ${\bm A}$ is the anti-symmetrization operator given by,
\begin{equation}
{\bm A}=\frac{1}{10}(1-9P^{OSTC}_{36})
\end{equation}
where $P^{OSTC}_{36}$ is the permutation operator for the quarks $3$ and $6$ and $OSTC$ stands for orbital, spin, isospin and color. Thus $P^{OSTC}_{36}$ operator exchanges the orbital, spin, isospin and color quantum numbers of the quarks $3$ and $6$.

The anti-symmetrization operator splits each term in the Hamiltonian into two parts: direct part and the exchange part. The direct part corresponds to the self energy of the nucleons and the exchange part gives the interaction between the nucleons. At asymptotic distances, the exchange part of the interaction vanishes since the overlap of wave functions is absent.

The harmonic oscillator wave function used here is,
\begin{equation}
\phi({\bm r}_{i})=\frac{1}{(\pi b^{2})^{3/4}}exp(-\frac{1}{2b^{2}}({\bm r}_{i}-\frac{{\bm s}_{I}}{2})^{2})
\end{equation}
where $b$ is the oscillator size parameter and $s_{I}$ is the generator coordinate. 

The energy is then given by,
\begin{equation}
E=\frac{\langle \psi |H{\bm A}|\psi\rangle_{l}}{\langle \psi |{\bm A}|\psi\rangle_{l}}
\end{equation}
where the subscript $l$ indicates that the quantities have been projected to the angular momentum value $l$.

Here, each nucleon is considered as a cluster of three quarks and the two nucleon system is considered as cluster $A$ and cluster $B$. In RGM, the total wave function is expressed as an anti-symmetric product of the single particle wave functions. The total wave function of the six quark system is,
\begin{equation}
\psi_{TOT}(\xi_{A},\xi_{B},{\bm R}_{AB})={\bm A}[\phi_{A}(\xi_{A})\phi_{B}(\xi_{B})\chi({\bm R}_{AB})]
\end{equation}
where $\phi_{A}$ and $\phi_{B}$ are the internal wave functions of the individual clusters $A$ and $B$ respectively and $\chi$ is the relative wave function between the two clusters and $\bm A$ is the total anti-symmetric operator of the six quark system. To separate the total wave function given in equation (11), the following choice of coordinate is made,
\begin{eqnarray}
\nonumber \xi_{1}={\bm r}_{1}-{\bm r}_{2}, \xi_{2}={\bm r}_{3}-\frac{{\bm r}_{1}+{\bm r}_{2}}{2}, {\bm R}_{A}=\frac{1}{3}({\bm r}_{1}+{\bm r}_{2}+{\bm r}_{3}),\\
\nonumber \xi_{3}={\bm r}_{4}-{\bm r}_{5}, \xi_{4}={\bm r}_{6}-\frac{{\bm r}_{4}+{\bm r}_{5}}{2}, {\bm R}_{B}=\frac{1}{3}({\bm r}_{4}+{\bm r}_{5}+{\bm r}_{6}),\\
\nonumber {\bm R}_{AB}={\bm R}_{A}-{\bm R}_{B}, {\bm R}_{G}=\frac{1}{2}({\bm R}_{A}+{\bm R}_{B}).
\end{eqnarray}
Here ${\bm r}_{i}$ is the  coordinate of the $i^{th}$ quark, the coordinates $\xi_{A}=(\xi_{1},\xi_{2})$ and $\xi_{A}=(\xi_{3},\xi_{4})$ are the internal coordinates of the two clusters $A$ and $B$ respectively, ${\bm R}_{AB}$ is the relative coordinate between the two clusters and ${\bm R}_{G}$ is the center of mass coordinate of the total system.

Since the Hamiltonian is translationally invariant, $L^{l}_{ij}$ can be written as,
\begin{eqnarray}
\nonumber L^{l}_{ij}=\int[\phi^{+SM}_{A}({\bm r}_{1},{\bm r}_{2},{\bm r}_{3},\frac{{\bm s}_{I}}{2})\phi^{+SM}_{B}({\bm r}_{4},{\bm r}_{5},{\bm r}_{6},\frac{-{\bm s}_{I}}{2})Y^{*}_{lm}({\hat{\bm s}}_{I})][H-E]A\\
\nonumber [\phi^{SM}_{A}({\bm r}_{1},{\bm r}_{2},{\bm r}_{3},\frac{{\bm s}_{J}}{2})\phi^{SM}_{B}({\bm r}_{4},{\bm r}_{5},{\bm r}_{6},\frac{-{\bm s}_{J}}{2})Y_{lm}({\hat{\bm s}}_{J})]\prod_{k=1}^{6}d^{3}{\bm r}_{k}d{\hat{\bm s}}_{I}d{\hat{\bm s}}_{J}
\end{eqnarray}

To take into account all possible interactions between the quarks, we have to consider seven different types of operators for the potential $V_{ij}$ in the Hamiltonian. They are $V_{12DR}$, $V_{36DR}$, $V_{12EX}$, $V_{13EX}$, $V_{16EX}$, $V_{14EX}$ and $V_{36EX}$, where $DR$ stands for direct part of the quark interaction between quarks $i$ and $j$ and $EX$ stands for the corresponding exchange part.

\section{Results and Discussions}

To analyze the contributions of the various components of the Hamiltonian, we have plotted the diagonal elements of the various kernels of the singlet and triplet NN potentials as a function of the relative distance between the nucleons $(s_{I})$. The adiabatic potential is calculated using the Born-Oppenheimer approximation given by,

\begin{equation}
\nonumber V^{Ad}_{12}=\langle\psi_{1}(s_{I})|H|\psi_{2}(s_{I})\rangle-\langle\psi_{1}(\infty)|H|\psi_{2}(\infty)\rangle
\end{equation}

\begin{table}
\centering     
\begin{tabular}{ll}
\hline
\hline
M (MeV) & 160.6\\
E (MeV) & 40.0\\
$\alpha^{2}$ (MeV fm$^{2}$) & 200.01\\
b (fm) & 0.6\\
$\alpha_{s}$ & 6.5\\
c (fm$^{-1}$) & 0.3\\
$\alpha_{1}$ & 1.035994\\
$\alpha_{2}$ (MeV) & 2.016150\\
c$_{0}$ (fm$^{-1}$) & 1.7324\\
$m_{\pi}$ (MeV) & 140.0\\
\noalign{\smallskip}\hline
\end{tabular}
\caption{List of Parameters}
\label{tab:1} 
\end{table}

There are ten parameters in our model - the masses of the quarks ($M$), the confinement strength($\alpha^{2}$), the harmonic oscillator size parameter ($b$), the quark-gluon coupling constant ($\alpha_{s}$), the current confinement model parameter ($c$) and mass of the pion ($m_{\pi}$). The coupling constant $\alpha_{s}$ is fixed by the N-$\Delta$ mass splitting which comes from the color magnetic term of COGEP. We have chosen the value of oscillator size parameter to be 0.6 fm which is consistent with the experimental results of the charge distribution of the nucleons and the axial charge distribution \cite{17}. The fixing of the other parameters are discussed in \cite{15,16}. The parameters used are listed in Table 1.

\begin{figure*}
\centering
  \includegraphics[scale=0.9]{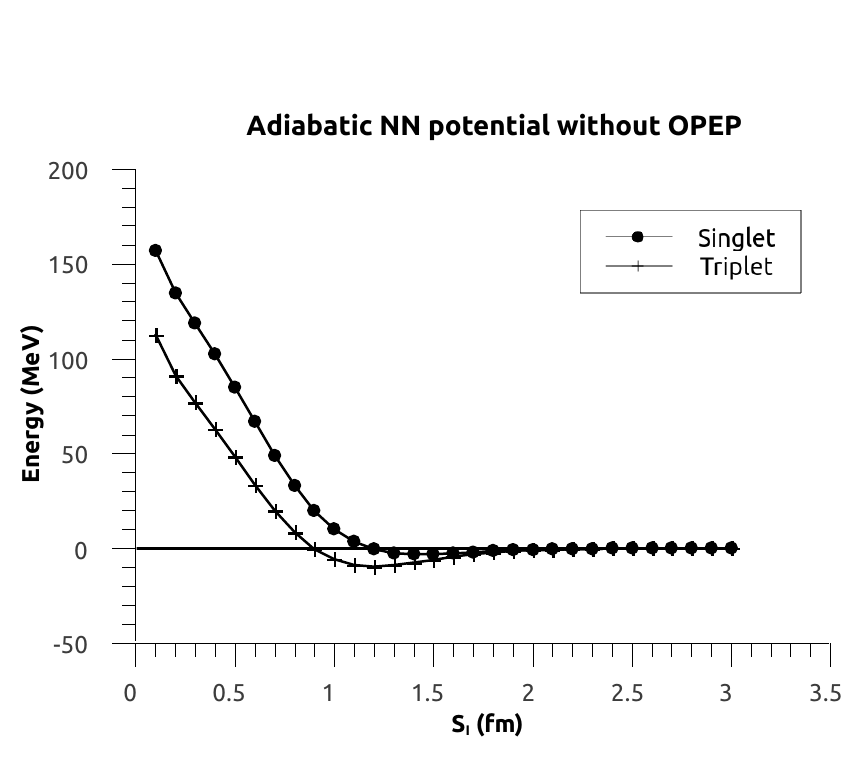}
\caption{Adiabatic NN potential without OPEP}
\label{fig:1}      
\end{figure*}

\begin{figure}[t]
\centering
\subfloat{\includegraphics[scale=0.8]{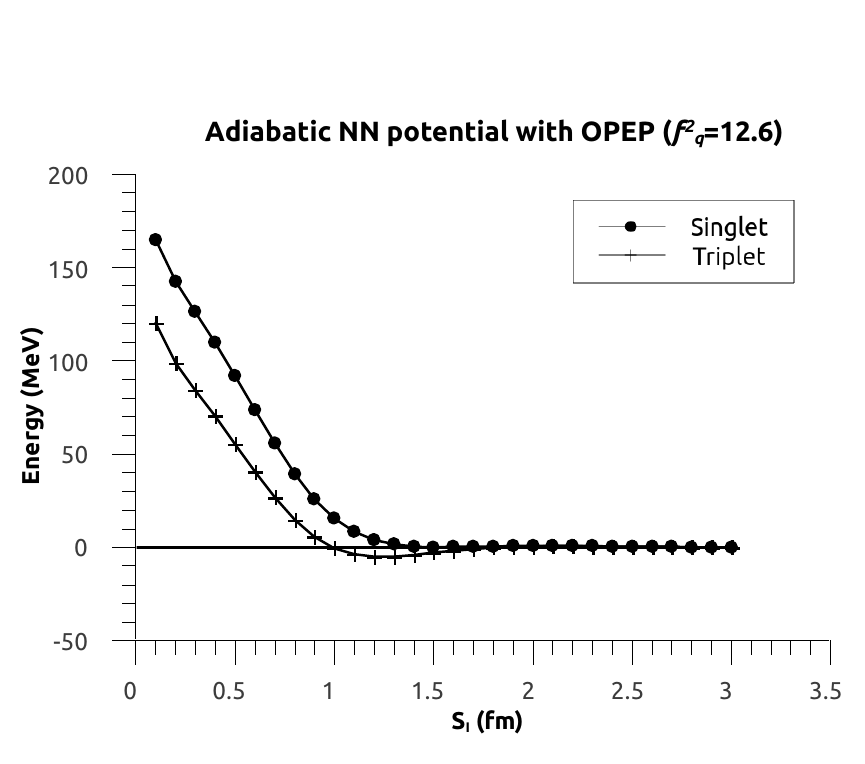}}
\subfloat{\includegraphics[scale=0.8]{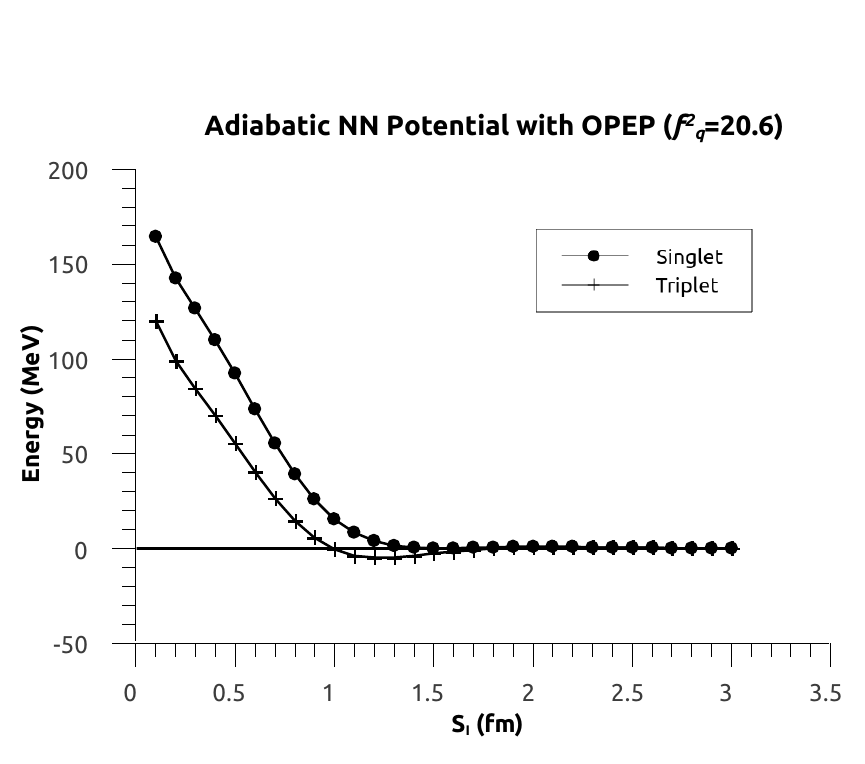}}
\caption{Adiabatic NN potential with OPEP; $f_{q}^{2}=12.6$ (left), $f_{q}^{2}=20.6$ (right)}
\label{fig:2}
\end{figure}

Figure \ref{fig:1} gives the plot of adiabatic NN potential without OPEP and Figure \ref{fig:2} gives the plot of adiabatic NN potential with OPEP. The intermediate range attraction for $^{1}S_{0}$ state is reduced in the presence of OPEP. The change in the potential with change in quark pion coupling constant value $f_{q}^{2}$ can also be seen from Figure \ref{fig:2}. The short range repulsion for $^{1}S_{0}$ state is larger than that of $^{3}S_{1}$ state. The color magnetic part of the COGEP can be considered as the reason for this difference.

\begin{figure*}
\centering
  \includegraphics[scale=0.9]{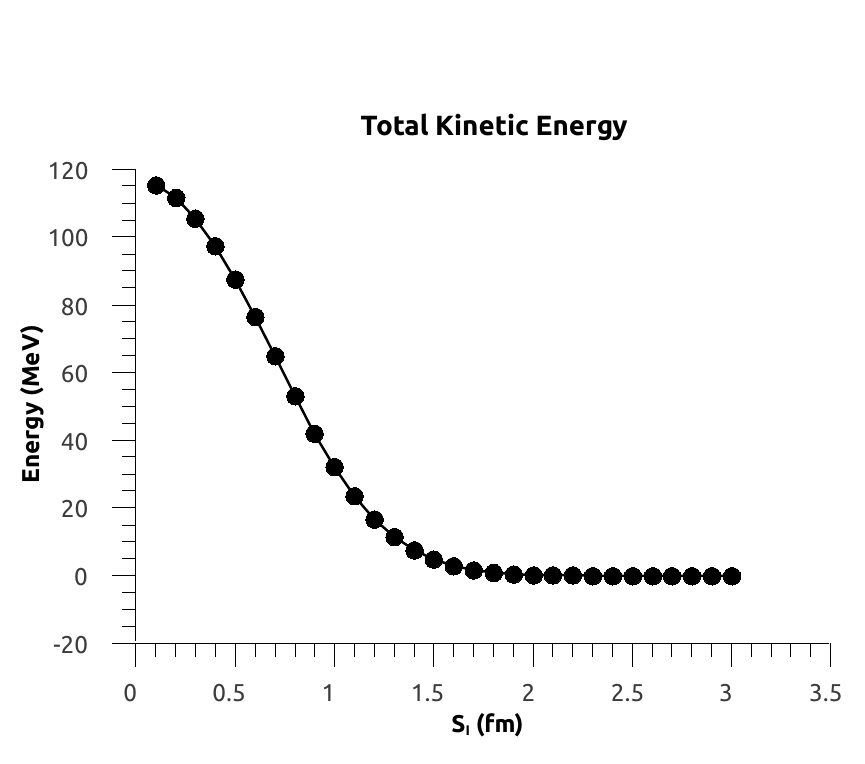}
\caption{Kinetic energy}
\label{fig:3}      
\end{figure*}

Figure \ref{fig:3} gives the plot of adiabatic kinetic energy vs $s_{I}$ (fm). It is clear from the figure that there is substantial contribution to the short range repulsion from the kinetic energy. Since direct part of kinetic energy do not contribute to the NN interaction the whole of the repulsion arises from the exchange part of the kinetic energy.

At short distances, the exchange kernels of $\delta^{3}(\bm r)$ is dominant over the exchange kernels of $c^{2}D_{0}(\bm r)$ and hence provide the short range repulsion. The short range repulsion is larger for $^{1}S_{0}$ state than the $^{3}S_{1}$ state. In the intermediate and long ranges, the exchange kernels of $c^{2}D_{0}(\bm r)$ dominates over the exchange kernels of $\delta^{3}(\bm r)$ and hence provide the intermediate and long range attraction. The intermediate range attraction is larger for the $^{3}S_{1}$ state than the $^{1}S_{0}$.

\begin{figure}[t]
\centering
\subfloat{\includegraphics[scale=0.8]{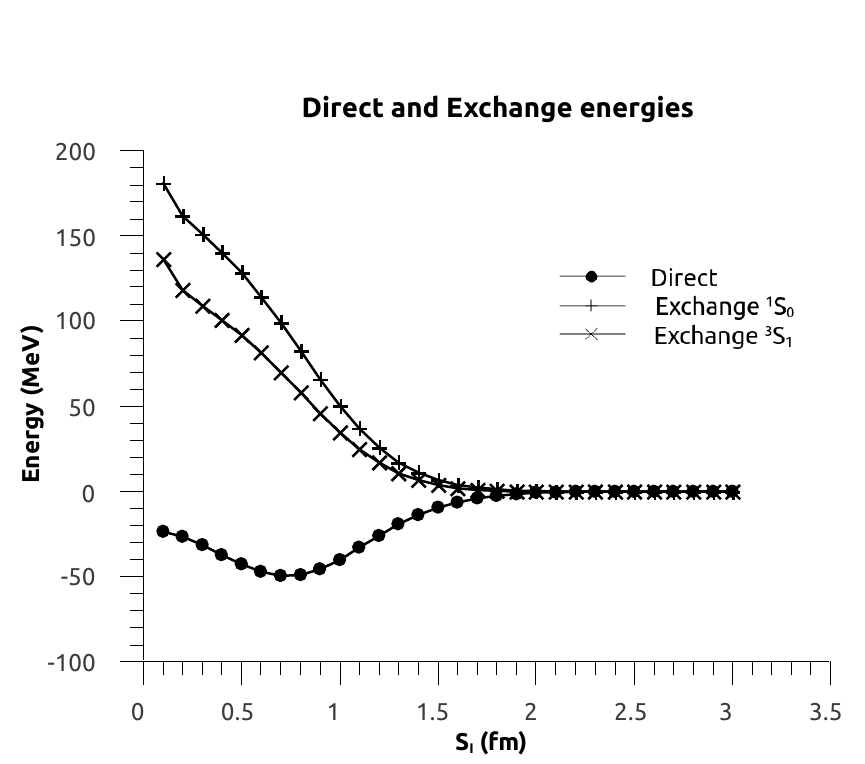}}
\subfloat{\includegraphics[scale=0.8]{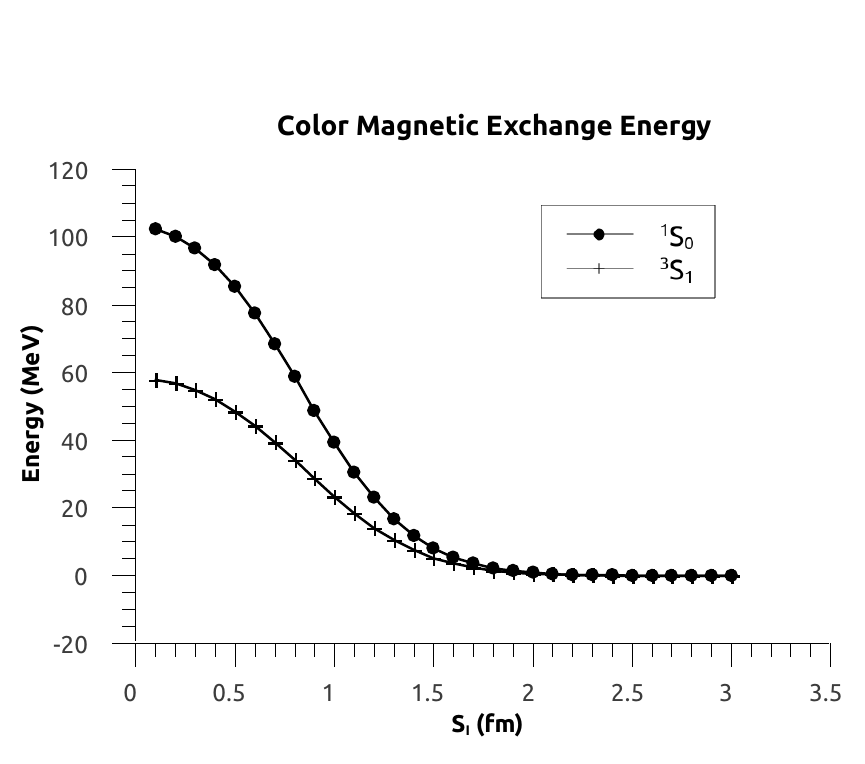}}
\caption{Direct and exchange components of NN potential (left) and color magnetic exchange part of COGEP (right).}
\label{fig:4}
\end{figure}

Figure \ref{fig:4} is a plot of the direct and exchange parts of the Hamiltonian and the exchange part of the color magnetic interaction in the adiabatic limit. The exchange part of the potentials of $^{1}S_{0}$ and $^{3}S_{1}$ states show repulsion in the short range. The $^{1}S_{0}$ state exchange potential is completely repulsive and that of the $^{3}S_{1}$ state shows a small attraction in the intermediate range. There is consistency with established results in the case of repulsive contribution to the adiabatic potential at short range, both to $^{1}S_{0}$ and $^{3}S_{1}$ states \cite{17,30,31}. There is no contribution from the color electric term to the NN interaction. Since the energy difference between $2(0s)^{3}$ and $(0s)^{6}$ configuration must come from the expectation value of ${\bm \lambda}_{i}.{\bm \lambda}_{j}$, the radial matrix elements are same for $2(0s)^{3}$ configuration and $(0s)^{6}$ configuration. The expectation value of the ${\bm \lambda}_{i}.{\bm \lambda}_{j}$ depends only on the number of quarks and hence the color electric elements of the COGEP and the confinement potential do not contribute NN adiabatic potential. The expectation value of ${\bm \lambda}_{i}.{\bm \lambda}_{j}$ ${\bm \sigma}_{i}.{\bm \sigma}_{j}$ in the color magnetic part of the COGEP for the $2(0s)^{3}$ and $(0s)^{6}$ configuration does not vanish and thus, the color magnetic part provides short range repulsion \cite{17}.

\section{Conclusions}

In this work, we have investigated the role played by the COGEP and OPEP on $^{1}S_{0}$ and $^{3}S_{1}$ NN adiabatic potentials using the Born-Oppenheimer approximation without evoking $\sigma$ meson in the framework of the RHM and the CCM by employing the RGM formalism. 

The aim was to understand the role played by the confined gluon exchange and one pion exchange on NN interaction. The COGEP was derived using the CGP. The major conclusions drawn from the present work are:
\begin{enumerate}
\item
The term arising out of the confinement of gluons is responsible for intermediate range attraction and hence plays the role of $\sigma$ mesons. The color magnetic terms of the COGEP yield both short range repulsion and intermediate range attraction for $^{1}S_{0}$ and $^{3}S_{1}$ states. The calculation clearly demonstrates the importance of the confinement of gluons on $^{1}S_{0}$ and $^{3}S_{1}$ state NN potentials.
\item
The OPEP and exchange terms of Kinetic energy provides state independent repulsion. 
\end{enumerate}

To conclude, we have obtained $^{1}S_{0}$ and $^{3}S_{1}$ NN adiabatic potentials starting from the dynamics of quarks and gluons including confinement of color and OPEP as a phenomenological input. Further study is in progress.

\end{document}